\documentclass[preprint2]{aastex}
\usepackage[dvips]{color} 

\shorttitle{Exploring the central sub-pc region of 3C~84}
\shortauthors{Suzuki, Nagai, \& Kino et al.}

\begin{document}


\title{Exploring the central sub-pc region of the $\gamma$-ray bright radio galaxy 3C~84 with the VLBA at 43~GHz in the period of 2002-2008}

\author{Kenta Suzuki,\altaffilmark{1} Hiroshi Nagai,\altaffilmark{2} Motoki Kino,\altaffilmark{2} Jun Kataoka,\altaffilmark{3} Keiichi Asada,\altaffilmark{4} Akihiro Doi,\altaffilmark{5} Makoto Inoue,\altaffilmark{4}  Monica Orienti,\altaffilmark{6, 7} Gabriele Giovannini,\altaffilmark{6} Marcello Giroletti,\altaffilmark{6} Anne L\"ahteenm\"aki,\altaffilmark{8} Merja Tornikoski,\altaffilmark{8} Jonathan Le\'on-Tavares,\altaffilmark{8} Uwe Bach,\altaffilmark{9} Seiji Kameno,\altaffilmark{10} Hideyuki Kobayashi,\altaffilmark{2}
}

%% are identified by the \altaffilmark after each name. Specify alternate

\altaffiltext{1}{Institute of Astronomy, University of Tokyo, 2-21-1 Osawa, Mitaka, Tokyo 181-0015, Japan}
\altaffiltext{2}{National Astronomical observatory of Japan, 2-21-1 Osawa, Mitaka, Tokyo 181-8588, Japan}
\altaffiltext{3}{Research Institute for Science and Engineering, Waseda University, 3-4-1, Okubo, Shinjuku, Tokyo, 169-8555, Japan}
\altaffiltext{4}{Institute of Astronomy and Astrophysics, Academia Sinica. P.O. Box 23-141, Taipei 10617, Taiwan, R.O.C.}
\altaffiltext{5}{Institute of Space and Astronautical Science, Japan Aerospace Exploration Agency, Yoshinodai 3-1-1, Chuo-ku, Sagamihara 252-5210, Japan}
\altaffiltext{6}{INAF Istituto di Radioastronomia, via Gobetti 101, 40129, Bologna, Italy}
\altaffiltext{7}{Dipartimento di Astronomia, Universita' di Bologna, via Ranzani 1, I-40127, Bologna, Italy}
\altaffiltext{8}{Aalto University  Mets\"{a}hovi Radio Observatory,  Mets\"{a}hovintie 114, FIN-02540 Kylm\"al\"a, Finland}
\altaffiltext{9}{Max-Planck-Institut f\"{u}r Radioastronomie, Auf dem H\"{u}gel 69, 53121 Bonn, Germany}
\altaffiltext{10}{Faculty of Science, Kagoshima University, 1-21-35 Korimoto, Kagoshima 890-0065, Japan}


\begin{abstract}
Following the discovery of a new radio component right before the GeV $\gamma$-ray detection since 2008 August by {\it Fermi} Gamma-ray Space Telescope, we present a detailed study of the kinematics and lightcurve on the central sub-pc scale of 3C~84 using the archival VLBA 43-GHz data covering the period between 2002 January to 2008 November. We find that the new component ``C3'', previously reported by the observations with the VLBI Exploration of Radio Astrometry (VERA), was already formed in 2003. The flux density of C3 increases moderately until 2008, and then it becomes brighter rapidly after 2008. The radio core, C1, also shows a similar trend. The apparent speed of C3 with reference to the core C1 shows moderate acceleration from $0.10c$ to $0.47c$ between 2003 November to 2008 November, but is still sub-relativistic. We further try to fit the observed broadband spectrum by the one-zone synchrotron self-Compton (SSC) model using the measured apparent speed of C3. The fit can reproduce the observed $\gamma$-ray emission, but does not agree with the observed radio spectral index between 22 and 43~GHz.
 
%%%%%%%%%%%%%%%%%%%%%%%
%We can see the feature of deceleration of C3, indicating deceleration jet. This deceleration view is feasible for results of VERA and this study.
%Additionally, relative position angle of C3 measured from pc scale core C1 is differ from that of former jet component C2 by $\sim $ 40$^{\circ}$. This suggests that advancing direction of C3 changed drastically within a few years. This raises an difficult problem on mechanism for jet direction.
\end{abstract}


\keywords{galaxies: active --- galaxies: individual(NGC~1275, Perseus A, 3C~84) --- galaxies: jets --- gamma rays: galaxies}

%% and \citet commands to identify citations. The citations are

%% their objects with \objectname{} or \object{}. Each macro takes the
%% differs from what is to be printed in the paper. The text appearing 
%% Otherwise, LaTeX will issue an error. 

\section{INTRODUCTION}\label{sec:intro}

The radio source 3C~84 is associated with the giant elliptical galaxy NGC~1275 ($z=0.0176$). It is well known that 3C~84 has a pair of compact radio lobe structures on the central 10-pc scale \citep{1994ApJ...430L..41V, 1994ApJ...430L..45W, 2000ApJ...530..233W, 2006PASJ...58..261A}. Because of its proximity, it allows us to study the region within the central sub-pc scale using a Very Long Baseline Interferometer (VLBI). It is an ideal laboratory to investigate the formation mechanism of relativistic jet ultimately powered by super-massive black holes and the interaction between the jets and ambient matter in the galactic central regions.

Thanks to the recent observations with {\it Fermi}/LAT, we have a new opportunity to explore GeV $\gamma$-ray production mechanism in misaligned radio-loud AGNs \citep{2010ApJ...720..912A}. With CGRO/EGRET, a few extragalactic radio galaxies, such as Centaurus A \citep{1999APh....11..221S}, 3C~111 \citep{2008ApJ...688..852H}, and NGC~6251 \citep{2002ApJ...574..693M} had been already detected. However, the detection of NGC~1275 \citep{2009ApJ...699...31A} by {\it Fermi}/LAT in 2008 August is particularly noteworthy because NGC~1275 was not detected by CGRO/EGRET \citep{2003ApJ...588..155R}. The flux density detected by {\it Fermi}/LAT is about 7 times higher than the upper limit of EGRET sensitivity. Intriguingly, the radio monitoring also shows the flux increase starting in 2005. The time variation of $\gamma$-ray flux density shows a similar trend with the radio flux density on the timescale of decades, implying the possible connection between the $\gamma$-ray emission and variable radio component.

Overall spectral energy distribution (SED) of NGC~1275 from radio to $\gamma$-ray can be explained by the synchrotron-self Compton (SSC) model and the deceleration jet model \citep{2009ApJ...699...31A} adopting sub-pc for the size of emitting region. The authors derived Lorentz factors of the emitting region  $\Gamma=1.8$ ($\theta=25$ deg, $\delta=2.3$) for the case of SSC model and $\Gamma$ varying from $10$ to $2$ ($\theta=20$ deg, $\delta$ varying 1.6 to 2.7) for the decelerating model, where $\theta$ and $\delta$ are the jet angle to the line of sight and the beaming factor of the emitting region, respectively. Here we should stress that the size of the $\gamma$-ray emitting region adopted in the SED models is comparable to a VLBI component size. Therefore, it is essential to test the scenario of co-spatiality of GeV $\gamma$-ray and radio emitting regions by VLBI observations.

In order to find the radio counterpart of the GeV $\gamma$-ray emitting region, we have conducted VLBI observations using the VLBI Exploration of Radio Astrometry (VERA) at 22~GHz between 2006 June 14 and 2009 April 24 (\citet{2010PASJ...62L..11N}, hereafter Paper I). Surprisingly, we found that the monotonic increase of radio flux density mainly originated in the newly born bright component C3  (Paper I). The measured projected speed of C3, however, was $(0.23 \pm 0.01)c$ on average between 2007 October 24 and 2009 April 24 in the central sub-pc to pc scale. Given previously estimated jet angle to the line of sight ($11^{\circ}$ - $55^{\circ}$: \citet{2006PASJ...58..261A, 2009AJ....138.1874L, 2009ApJ...699...31A}), the de-projected speed corresponds to 0.24$c$ - 0.55$c$, i.e. slower than the speed of jet derived from the SED modeling \citep{2009ApJ...699...31A}.
The result of VERA observation implies $\delta < 2$, and there seems to be a discrepancy between the SED models and VLBI observations.

A possible idea to explain this discrepancy is that a jet component with relativistic speed other than C3, located much closer to the central black hole (within the central sub-pc), is responsible for the GeV $\gamma $-ray emission. If such a relativistic motion in the jet of 3C~84 occurs only in vicinity of the black hole, the speed of C3 could have been relativistic soon after the ejection from the core, and then it underwent to a rapid deceleration. \citet{2010A&A...516A...1L} also suggested that jets are relativistic in bright cluster galaxies but they decelerate very soon (sub-parsec scale) because of a strong interaction with the ISM. In Paper I, we argued possible detection of the relativistic speed at the early stage of emergence of C3. However, this was not conclusive because of (1) the lack of spatial resolution of VERA at 22~GHz and (2) possible screening by free-free absorption.

In order to overcome the above potential problems, higher resolution observations at higher frequency are required. We explore the possible relativistic flow of C3 at the early stage of its emergence from the core with the archival VLBA data at 43~GHz. 
We selected only data obtained from 2002 to 2008, i.e. considering earlier epochs than those discussed in Paper I.
We also discuss the SED fit using one-zone SSC model involving the actual measured apparent motion of C3 and possible ideas for incorporating the VLBI observation with GeV  $\gamma $-ray emission.

Throughout this paper, we adopt the following cosmological parameters; $H_{0}=71$~km sec$^{-1}$ Mpc$^{-1}$, $\Omega_{\mathrm{M}}=0.27$, and $\Omega_{\mathrm{\Lambda}}=0.73$ (1~mas = 0.353~pc, and 0.1~mas~yr$^{-1}$ = $0.113c$). 

\section{OBSERVATION AND DATA ANALYSIS}\label{sec:obsanaly}

\subsection{The VLBA at 43~GHz}

First, we examined all of the (available) archival VLBA data at 43~GHz obtained between 2002 to 2008 from the data archive system of the National Radio Astronomy Observatory (NRAO). Data reduction was performed using Astronomical Image Processing System (AIPS). $A$ $priori$ amplitude calibration was done based on the measurements of the system noise temperature ($T_{\rm sys}$) during the observations and the elevation dependent antenna gain provided by each station. In this process, we also applied opacity correction due to the atmospheric attenuation, assuming that the time variation of the opacity is not significant during each observation. Fringe fitting was performed with the AIPS task FRING. After applying the delay and rate solutions, we combined each set of data into one channel across the 8-MHz bandwidth. Imaging and self-calibration process was performed using Difmap software package \citep{1994BAAS...26..987S}. The final images were produced after iterations of CLEAN, phase and amplitude self-calibration processes on successively shortened calibration time. Second, we excluded the entire datasets with relatively poor dynamic range (because of bad weather condition or lack of baselines). Finally, we ended up with 28-epoch datasets (shown in Table \ref{tab:obsinfo}). In almost all the archival data we analyzed, 3C~84 was observed as a calibrator and the typical observing time is less than 10 min.

\subsection{The Mets\"{a}hovi at 37~GHz}

The single dish monitoring data at 37~GHz is adopted from Mets\"{a}hovi quasar monitoring program \citep{1998A&AS..132..305T, 2011A&A...532A.146L}. The observations were carried out with Mets\"{a}hovi 14-m Radio Telescope. In order to compare the VLBI and single dish lightcurves, we selected the data from 2001 December to 2008 December. Details of these observations and calibration are described in \citet{1998A&AS..132..305T}.

\section{MODELING THE VLBA DATA}

\subsection{Gaussian Model Fittings}

To obtain the flux density of each component of 3C~84 jet, we modeled the source structure using the Difmap task modelfit. In the modelfit process, we fitted multiple elliptical Gaussian model components to the visibility data of each epoch. We adopted the fitting model components  that gave a good fit to the data as judged by relative $\chi^{2}$-squared statistics. In some cases, we adopted circular Gaussian models or point sources instead of the elliptical Gaussian if either of them were capable of reproducing a better fit. 
We consider the fit acceptable if the model components contain more than 95\% of the total CLEANed flux.
Resultant images mainly consist of three major components (C1, C2, and C3; shown in \S4), although C2 is not identified in some epochs because of low signal-to-noise ratio. In addition to the major components, some faint minor components are fitted. The choice of adding a few minor components to the model is somewhat arbitrary, and it is difficult to maintain the consistency across the epochs. We note, however, that the choice of minor components does not appreciably affect the fitting results of the major components.

In order to check that the fit is not local minimum, we confirmed that the result is unchanged significantly by choosing different starting Gaussian models.

\subsection{Positional Error of VLBA Images} \label{ssec:error}
In general, it is difficult to measure a positional error of each component quantitatively from single epoch interferometric data. In particular, a systematic error, which arises from the deconvolution error, the presence of other nearby jet features and so on, is difficult to estimate separately. Hence the sampled interferometric-visibility on the $(u, v)$-plane is poor, and therefore systematic error of component position may be large compared to long-track observations.  For these reasons, we estimated the component positional errors by examining the scatters in the positions of components (defined by the AIPS task MAXFIT) in images between two close epochs (within 10-day separation), such that the jet has approximately the same structure in both epochs. 
The apparent motion of $1c$ will show 2.4 $\mu $as/day positional change ($<$ 0.01 of typical VLBA 43-GHz beam), so the motion of the component is negligible within 10 days in this study. We have done this analysis on 11 pairs of images (Table \ref{tab:pairs}). 

\begin{figure}[hbt]
\includegraphics[clip,width=\linewidth]{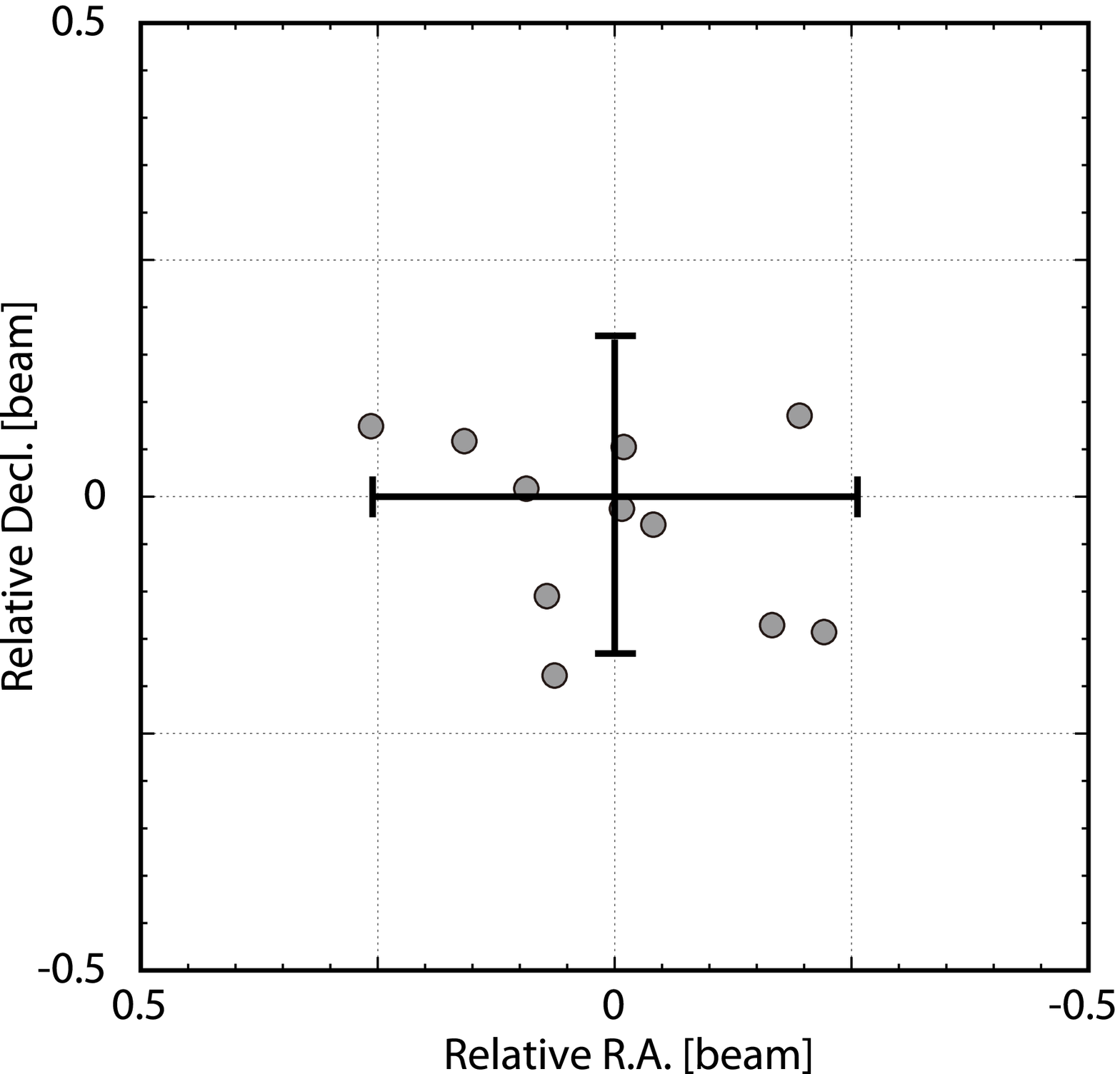}
\caption{
Estimation of position measurement accuracy for C3. Each point represents the difference of C3 position measured between two
epochs closely separated in time (within 10 days), which is normalized by the beam size averaged between the pair. Thick bars correspond
to the standard deviation of the scatter for right ascension and declination direction, and are estimated to be 0.255 $\theta_{\rm beam}$  and 0.166 $\theta_{\rm beam}$, respectively.\label{fig:err}
}
\end{figure}

Figure \ref{fig:err} shows the differences of position of C3 for these 11 pairs.  Each data point is normalized with the beam size averaged between the pair. The standard deviations along right ascension ($\sigma_{s}^{_{\rm RA}}$) and declination ($\sigma_{s}^{_{\rm DEC}}$) normalized by beamsize ($\theta_{\rm beam}$) are  0.150 and 0.098, respectively. As in \citet{1994saes.book.....B}, the $100\alpha$\% confidence interval of the standard deviation of population for statistical ensemble $i$, $\sigma_{p}^i$, is estimated from $N$ sample of statistical ensemble $i$ as 
\begin{equation}
\frac{N{\sigma_{s}^i}^{2}}{\chi_{N-1}^{2}(\alpha/2)} < \sigma_{p}^i < \frac{N{\sigma_{s}^i}^{2}}{\chi_{N-1}^{2}(1-\alpha/2)},
\end{equation}
where $\chi_{N}^{2}(\alpha)$ is $\chi ^2$ statistics for d.o.f. = $N$ on which the event $\chi ^2 > \chi_{N}^{2}$ occurs with probability $\alpha $. We apply this estimator for right ascension ensemble ($i$ = RA) and declination ensemble ($i$ = DEC). Given $N$ = 21 and $\alpha$ = 0.95, the standard deviation of population for both right ascension ($\sigma_{p}^{_{\rm RA}}$) and declination ($\sigma_{p}^{_{\rm DEC}}$) is
\begin{eqnarray}
0.106& < \sigma_{p}^{_{\rm RA}} < 0.255, \\
0.070& < \sigma_{p}^{_{\rm DEC}} < 0.166.
\end{eqnarray}
Hereafter, the positional error of each component is set as 0.255 $\theta_{\rm beam}$ for right ascension and 0.166 $\theta_{\rm beam}$ 
for declination, where $\theta_{\rm beam}$ is the beam size on each epoch.

\section{RESULTS}\label{sec:results}

\subsection{Overall Structures}\label{ssec:images}

Figure \ref{fig:3C84_Q_all} shows the total intensity images at 43~GHz. Information on all images is also summarized in Table \ref{tab:VLBA_Q_sum}. All images are convolved with the circular restoring beam with a diameter of 0.3~mas. After 2003 November 20, a new component C3 seems to emerge south of C1. This new component gradually becomes brighter with increasing separation from component C1.  Emergence of C3 is also seen in 22-GHz observations (Paper I) around 2007, but the VLBA 43-GHz observations find that the component has already emerged in earlier epoch, because of thin opacity and/or higher resolution than the VERA 22~GHz observations. After the emergence, C3 is advancing toward position angle $\sim $ 170$^\circ $ measured from C1. Its direction differs from that of C2 with respect to C1 by 40-50 degrees. No counter jet component is detected with a level of $2\sigma$ throughout all epochs. Physical parameters of all fitted components are listed in Tables \ref{tab:model_C1}, \ref{tab:model_C2}, and \ref{tab:model_C3}.

\subsection{Kinematics}\label{ssec:motion}

\begin{figure}[htb]
\includegraphics[clip,width=\linewidth]{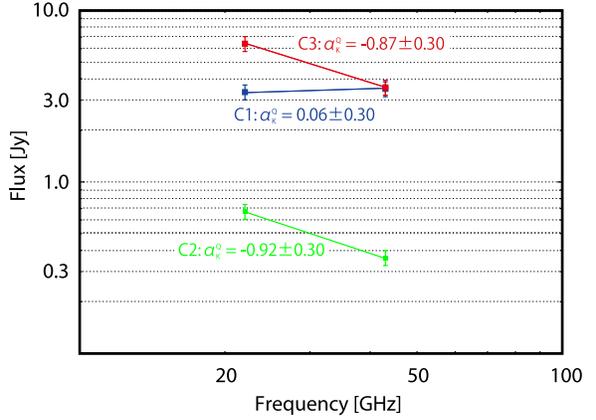}
\caption{
Spectral indices of C1, C2 and C3 between 22 GHz and 43 GHz at 2008 November 27. We assume that flux error is 10\% of the flux density of each component. Spectral indices of C1, C2, and C3 are 0.06 $\pm $ 0.30, -0.92 $\pm $ 0.30, and -0.87 $\pm $ 0.30, respectively. \label{fig:spectrum}
}
\end{figure}

We use the AIPS task MAXFIT to define the (relative) positions of components at each epoch. Since absolute position in each image is lost at the self calibration and fringe fitting processes, we need to define the reference position for the argument of kinematics. In this work, we will evaluate kinematics with reference to the optically-thick radio core. Thus we attempt to measure the spectral index of each component to define the reference position. We measured the spectral index $\alpha$ for each component at 2008 November 27, when 22~GHz and 43~GHz observations were carried out only one day apart. We define the spectral index between 43~GHz and 22~GHz as $\alpha _{_K}^{_Q} = \log (S_{Q}/S_{K})/\log (43/22)$, where $S_{Q}$ and $S_{K}$ are flux of components at 43~GHz and 22~GHz. As shown in Figure \ref{fig:spectrum}, the spectrum of C1 is flat ($\alpha _{_K}^{_Q} \sim 0$) while those of C2 and C3 are steep ($\alpha _{_K}^{_Q} \sim -0.9$). It seems natural to interpret C1 as the radio core and C2 and C3 as optically-thin jet components. Hereafter, we regard C1 as the reference position. 

\begin{figure}[htb]
\includegraphics[clip,width=\linewidth]{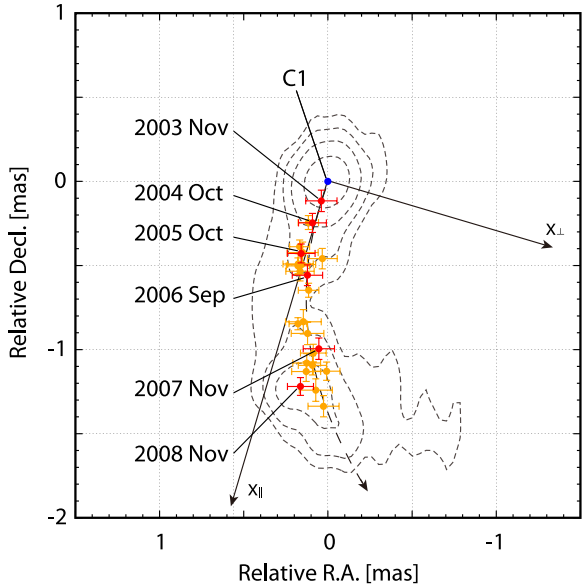}
\caption{
Peak position of C3 for all epochs, superposed on the contours of 43-GHz intensity distribution on 2008 November 27. The dashed line traces the average positional change of C3. We set $x_{\parallel }$ and $x_{\perp }$ coordinates to evaluate the motion parallel and perpendicular to the initial position angle of C3, 161.4$^{\circ }$ from C1.
\label{fig:positions}
}
\end{figure}

\begin{figure}[htb]
\includegraphics[clip,width=\linewidth]{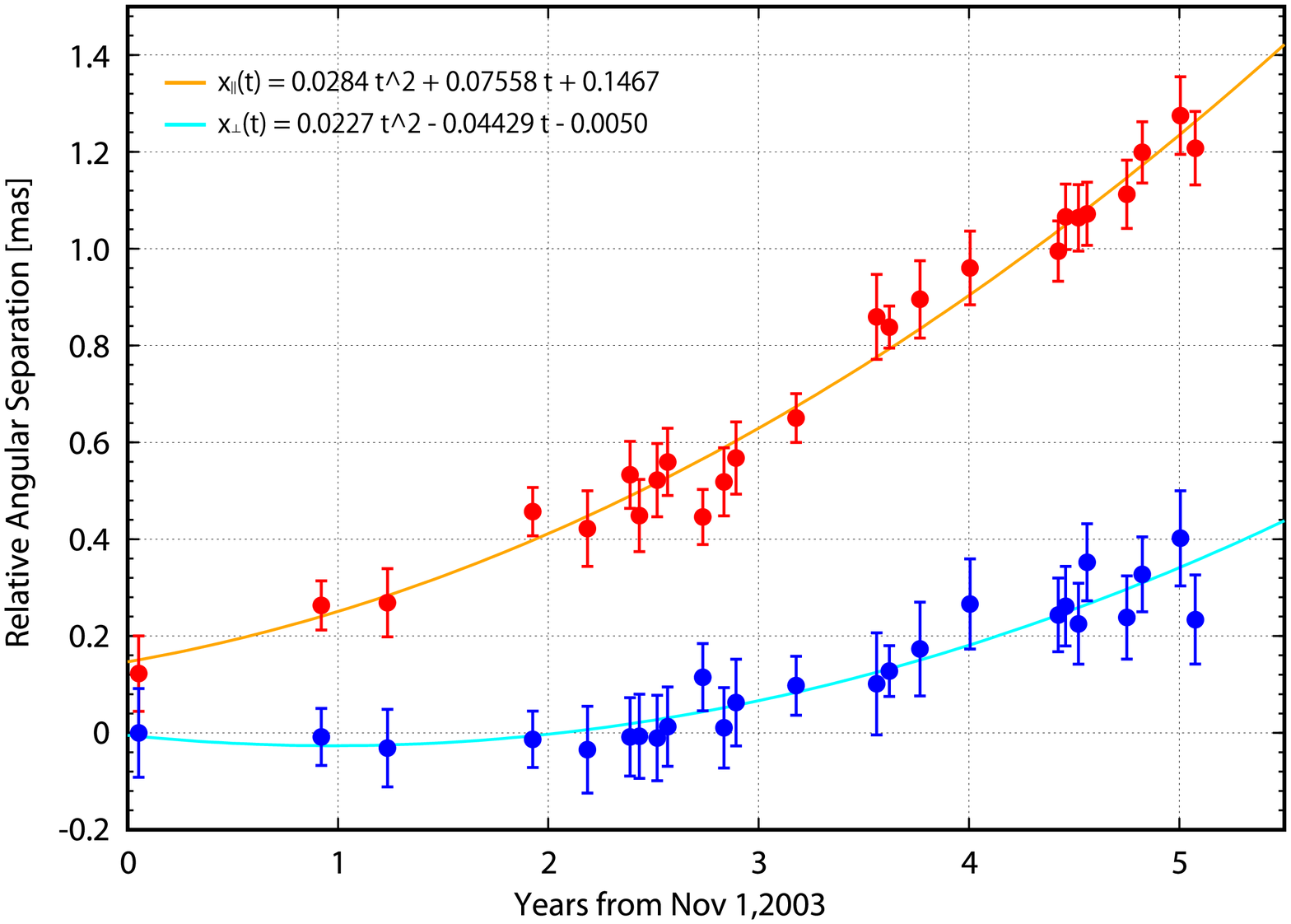}
\caption{
The change of relative angular distance between C1 and C3 in $x_{\parallel }$ and $x_{\perp }$ coordinates as a function of time from 2003 November 1 in years. Red and blue points show $x_{\parallel }$ and $x_{\perp }$ for each epoch. Orange and cyan lines are fitted functions. \label{fig:separation}
}
\end{figure}

\begin{figure}[htb]
\includegraphics[clip,width=\linewidth]{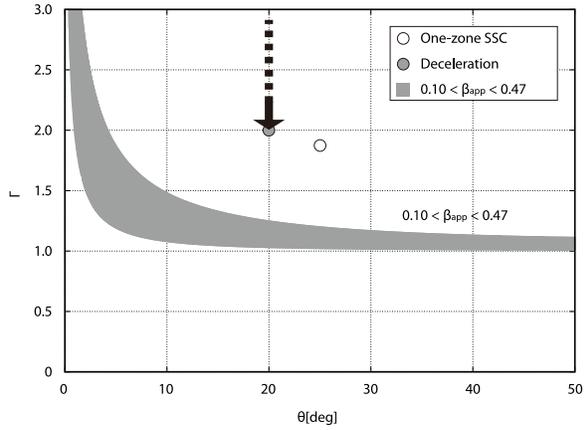}
\caption{
Constraint on the bulk Lorentz factor $\Gamma $ and the jet angle to the line of sight $\theta $. The gray-filled area (corresponding $0.10 < \beta _{\rm app} < 0.47$) is allowed for the apparent speed range of C3 between 2003 November and 2008 November. 
The white circle and gray-filled circle show the best-fit parameters for one-zone SSC model ($\Gamma = 1.8$ with $\theta =25^{\circ }$) and  for deceleration jet model (as shown by the arrow, jet decelerates from $\Gamma_{\rm max} = 10.0$ to $\Gamma_{\rm min} = 2.0$ over a distance of $5 \times 10^{17}$ cm (0.16 pc), with $\theta =20^{\circ }$) from \cite{2009ApJ...699...31A}.
 \label{fig:gamma}
}
\end{figure}

Figure \ref{fig:positions} shows the change of the peak position of C3. The motion of C3 is mostly in the north-south direction, but also shows the motion in the east-west direction. To describe the positional change of C3, firstly we set a coordinate as shown in Figure \ref{fig:separation}. We define $x_{\parallel }$  axis to be parallel to the line between the position of C1 and C3 in 2003 November 20 (position angle = 161.4$^{\circ }$) and $x_{\perp }$ axis to be perpendicular to $x_{\parallel }$.
Then, we fitted the positional change of C3 using a quadratic function in consideration for the change in speed for each axis (Figure \ref{fig:separation}). Position of C3 in $x_{\parallel }$ and $x_{\perp }$ coordinates is parameterized as,
\begin{eqnarray}
x_{\parallel }(t) &= a~t^2 + b~t + c, \\
x_{\perp }(t)     &= d~t^2 + e~t + f,
\end{eqnarray}
where the units of $x_{\parallel }$ and $x_{\perp }$ are in mas, $t$ is the time from 2003 November 1 in years. The best fit parameters are, $a = (0.028 \pm 0.006)$, $b = (0.076 \pm 0.030)$, $c = (0.147 \pm 0.046)$, $d = (0.023 \pm 0.005)$, $e = (0.044 \pm 0.028)$, and $f = (0.005 \pm 0.039)$. The apparent speed and acceleration can be calculated from these parameters. The apparent speed of C3 is changing from $0.09 \pm 0.04$ mas yr$^{-1}$ to $0.41 \pm 0.07$ mas yr$^{-1}$ (or $\beta _{\rm app}$ = $0.10 \pm 0.05$ to $0.47 \pm 0.08$) between 2003 November 20 and 2008 November 27 with acceleration rate of $0.07 \pm 0.01$ mas yr$^{-2}$ (or $\dot{\beta }_{\rm app} =  0.08 \pm 0.02 $ yr$^{-1}$). The averaged $\beta _{\rm app}$ of C3 from 2003 November 20 to 2007 November 2, when C3 is not identified in Paper I, is derived to be $(0.23 \pm 0.06)c$. Although Paper I discussed possible detection of relativistic flow with $\beta _{\rm app} \sim 2.3$ during the middle of 2007, 
 we cannot find such a fast motion at earlier epochs from high-resolution VLBA data. Thus, we can rule out the existence of relativistic flow with $\beta _{\rm app} \sim 2.3$ during 2006 - 2007.
In Figure \ref{fig:gamma}, the curved line shows the constraint on the jet speed and jet angle to the line of sight, based on the observed projected speed of C3 ($0.10c$ - $0.47c$), together with the jet speed and jet angle adopted in \citet{2009ApJ...699...31A}.
There is a clear discrepancy between the measured motion and predicted jet speed from both one-zone SSC model and the deceleration model.

The averaged apparent speed of C3 from 2007 November 2 to 2008 November 27 is derived as $(0.43 \pm 0.09)c$ from interpolation. This value is somewhat larger than the result of Paper I ($\beta _{\rm app} = 0.23 \pm 0.01$ on average between 2007 October 24 and 2009 April 24). This inconsistency might be caused by (1) the difference of coordinate setting and time-dependence of fitting function, and/or (2) under-estimation of positional error in Paper I. Focusing on the epoch overlapping with Paper I (2007-2008), the separation of C3 is systematically larger than that in Paper I by $\sim 0.1$~mas as if expected from the position shift of reference position (peak position of C1) due to the optical depth.

C2 is too faint to accurately determine its peak position, and higher dynamic range data are needed to identify and discuss it's kinematics.

\subsection{Lightcurves}\label{ssec:LC}

\begin{figure}[htb]
\includegraphics[clip,width=\linewidth]{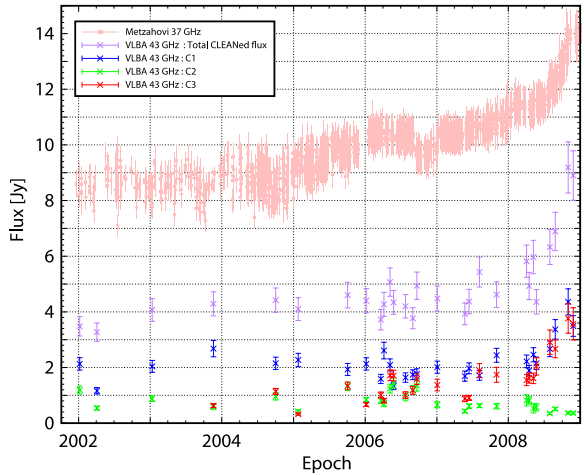}
\caption{
VLBI and single dish lightcurves of 3C~84. Pink and magenta points indicate the flux density of Mets\"{a}hovi
at 37~GHz \citep{2011A&A...532A.146L} and VLBA at 43~GHz, respectively. Flux densities of C1, C2, and C3 are indicated by blue, green, and red points, respectively.
\label{fig:LC}}
\end{figure}

\begin{figure}[htb]
\includegraphics[clip,width=\linewidth]{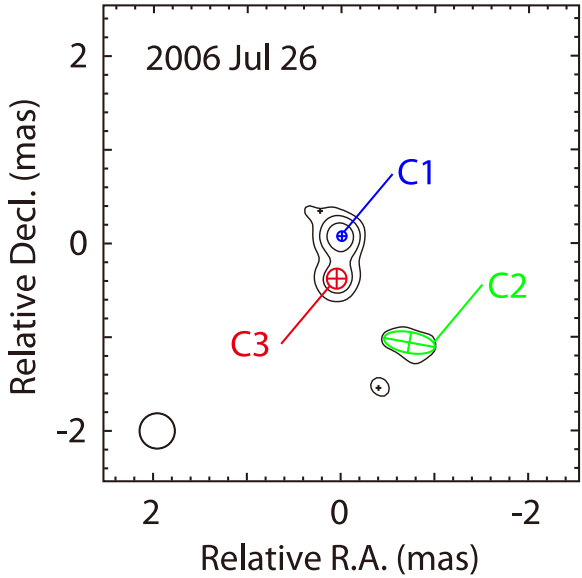}
\caption{
An example of model fitting result for VLBA 43-GHz image. Black contour shows the brightness distribution. Blue, green, and red symbols correspond to C1, C2, and C3, respectively. Black symbols are other minor components.\label{fig:model_ims}
}
\end{figure}

Figure \ref{fig:LC} shows the lightcurve of total CLEANed flux and the flux density of each component at 43~GHz, together with Mets\"{a}hovi 37-GHz lightcurve. 
The flux of each component is defined as the integrated flux of Gaussian component, which is identified with the Gaussian models obtained by the Difmap modelfit. An example of the result of Gaussian fitting is shown in Figure \ref{fig:model_ims}.
Overall trends of the 43-GHz VLBA total CLEANed flux and Mets\"{a}hovi 37-GHz lightcurve are very similar, and the flux density of Mets\"{a}hovi at 37~GHz is always brighter by $\sim $ 5 Jy than that of VLBA at 43~GHz throughout all epochs. This suggests that the flux variability essentially arises from the central pc region. 

From 2002 to 2008, the lightcurve of total VLBA flux density shows moderate increase. After 2008, rapid flux increase is clearly seen. C3 shows a significant flux density increase, which is consistent with Paper I. C1 also shows a flux density increase, although not obvious in Paper I. This could be because of large optical depth at 22~GHz. The increasing tendencies of C1 and C3 flux density resemble that of total CLEANed flux. This indicates that the radio flare in central $\sim 1$ pc region originates in C1 and C3, invoking the possibility that the GeV $\gamma $-ray emission originates in these innermost components. This view is consistent with the $\gamma $-ray emitting zone derived by \citet{2011A&A...532A.146L} (distance from the core $<$ 1pc). 

\section{SED MODELING AND INTERPRETATION}\label{sec:discussion}

\begin{figure}[htb]
\includegraphics[clip,width=\linewidth]{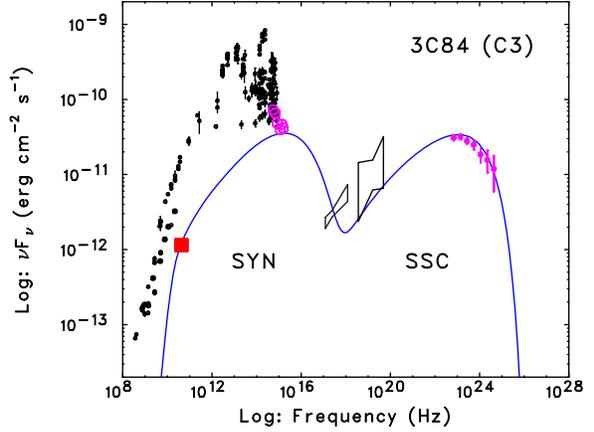}
\caption{Overall SED of NGC~1275 constructed with multiband data (black and magenta; \citet{2009ApJ...699...31A}) and the VLBA 43-GHz flux of C3 (red square; this work). The SED is fitted to VLBA 43-GHz flux of C3 (red square), optical/UV (Swift/UVOT; magenta open circle), and $\gamma $-ray ({\it Fermi}; magenta point) data with a one-zone SSC model. We obtained the model parameters as $R=1.45 \times 10^{17}~{\rm cm}~{\rm(0.133 mas)}$, $B=0.37~{\rm G}$, $q_{e}=5 \times 10^{-7}~{\rm cm^{-3}s^{-1}}$, $s=1.2$, $\gamma_{\rm min}=1$, $\gamma_{\rm max}=7.5\times 10^{4}$, and $\delta=1.65$, where $R$ is the comoving radius of the jet emitting region, $B$ is the mean magnetic field in the radiating plasma, $q_{e}$, $s$, $\gamma_{\rm min}$, $\gamma_{\rm max}$ are the normalization factor, the power-law index, maximum and minimum Lorentz factor for injection spectrum in \citet{2002ApJ...564...97K}. Doppler factor $\delta$ is derived from the measured value of $\beta _{\rm app} = 0.44$ for C3 at 2008 August and the assumed viewing angle $\theta $ = 25 deg (yielding $\Gamma $ = 1.19). The flux measured at 43~GHz on 2008 August 27th is $2.67\pm0.32~{\rm Jy}$. \label{fig:SED}
}
\end{figure}

In this section, we try to reproduce the broadband flux from $\gamma$-ray emitting region of 3C~84 based on the quantities of C3 obtained as in \S \ref{sec:results}. In Figure \ref{fig:SED}, we show the observed SED of 3C~84 and one-zone SSC model fit to the observed SED using the size, flux, and $\beta_{\rm app}$ of C3 component measured by the VLBA observations. The flux data are the same as the ones in \citet{2009ApJ...699...31A} but we add the 43-GHz flux of C3 measured on 2008 August 27. We assume that $\gamma$-ray and VLBA 43-GHz flux are emitted from the same sub-pc region. Optical data (after the extraction of host galaxy contribution; c.f., \citet{2009ApJ...699...31A}) is also used as flux upper limit of C3. The $\beta _{\rm app}$ at 2008 August is $0.44$ (in Figure \ref{fig:gamma}). Assuming the jet viewing angle as $25^{\circ}$ (\citet{2009ApJ...699...31A}), the intrinsic jet speed $\beta$ can be estimated as $0.54$, which corresponds to $\Gamma=1.19$ ($\delta=1.65$). The Gaussian-fitted model size of C3 is $\sim 0.2$-$0.3~{\rm mas}$. To fit the 43-GHz flux, the size of $\gamma$-ray emitting region is adopted as $1.45 \times 10^{17}~{\rm cm}~{\rm(0.133~mas)}$ which is slightly smaller than the C3 model size. The detailed numerical treatment of synchrotron and inverse Compton scattering processes is shown in \citet{2002ApJ...564...97K} and references therein.

If we adopt $\gamma_{\rm min} \sim 10^{3}$ as is the case in \citet{2009ApJ...699...31A}, 43~GHz lies below the frequency of low energy cutoff and therefore the spectral index should be $\alpha _{_K}^{_Q} = 1/3$ ($\nu f_{\nu} \propto 4/3$). This is inconsistent with the observed spectral index between 22 and 43~GHz $\alpha _{_K}^{_Q} = -0.87\pm0.3$ as shown in Figure \ref{fig:spectrum}. Then, we adopt $\gamma_{\rm min} = 1$ to raise the cutoff frequency. This fit is indicated by a solid line in Figure \ref{fig:SED}. The model spectrum requires an extremely hard electron energy index $s=1.2$ occasionally seen in TeV blazars (e.g., \citet{2000ApJ...528..243K}) if we try to reproduce the flux data at other wavelengths. The synchrotron opacity becomes unity at $\sim 30$~GHz, and the resultant model spectrum shows $\alpha _{_K}^{_Q} \sim -0.05$, which is still inconsistent with the observed spectral index. Hence it seems difficult to attribute the broadband SED to the one-zone SSC emission from C3. This indicates that some other components C1 and/or C2 also should be taken in account for the candidates for $\gamma$-ray emitting region.

\section{SUMMARY}\label{sec:summary}

\begin{figure}[h!]
\centering
\includegraphics[clip,width=\linewidth]{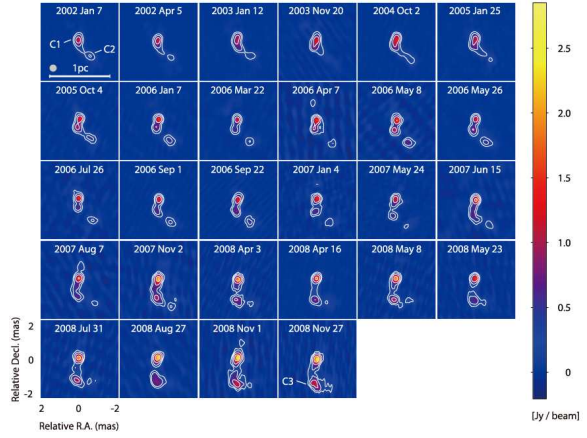}
\caption{
The VLBA 43-GHz images from 2002 January 7 to 2008 November 27. Image sizes are 4 $\times$ 4 mas$^2$ (1 mas corresponds to
0.353 pc). All images are convolved with the same restoring beam of 0.3 mas in diameter. Contour levels are plotted at the level of 68.4 $\times$ $n^2$
($n = 1, 2, 3, 4, $) mJy beam$^{-1}$. The lowest contour is the highest image noise r.m.s. among all images. \label{fig:3C84_Q_all}
}
\end{figure}

Following the detection of 3C~84 by {\it Fermi}/LAT in 2008, we discussed the kinematics of the newly formed jet component C3 using our VERA radio observations for the period 2006 June - 2009 April in Paper I. In this paper, we further explored the kinematics and lightcurves of C1, C2, and C3 using archival 43-GHz VLBA data spanning a time interval from 2002 January to 2008 November, i.e. before the VERA observations. Summary and discussions are as follows:

\begin{enumerate}
\item In the multi-epoch high resolution images obtained by VLBA at 43~GHz (Figure \ref{fig:3C84_Q_all}), we find that C3 has been visible since 2003 November. We succeed in obtaining a strong constraint on the time of the C3 birth before 2003 November. 

\item The lightcurve of VLBA at 43~GHz shows a trend similar to that of Mets\"{a}hovi at 37~GHz, indicating that the increase of radio flux density arises from the region within the central $\sim1$~pc.  In particular, the rapid increase of flux density has been seen since the middle of 2008. Both C1 and C3 show the largest change of flux density for this flare.  This indicates that the radio flare in the central $\sim1$~pc region originates in both C1 and C3.  The flux density increase of C1 was not prominent in Paper I, possibly because of large optical depth at 22~GHz.

\item The apparent speed of C3 with respect to C1 changes from $(0.10 \pm 0.05)c$ to $(0.47 \pm 0.08)c$ between 2003 November 20 and 2008 November 27 with acceleration rate $\dot{\beta }_{\rm app} = 0.08 \pm 0.02 $ yr$^{-1}$. No superluminal motion is detected with VLBA observations at 43~GHz before 2008 November. 

\item The one-zone SSC model using measured apparent speed of C3 is fitted to the observed broadband spectrum, in particular, the 43-GHz and GeV $\gamma $-ray fluxes. Our model fit is difficult to reproduce the optically-thin radio spectrum $\alpha _{_K}^{_Q} \sim -0.9$ that is measured by the VLBA observations. This indicates that it seems difficult to attribute the broadband SED to the one-zone SSC emission from C3. Therefore, the other components should not be excluded as candidates for $\gamma$-ray emitting regions.

\end{enumerate}

\acknowledgments

We thank anonymous referee for helpful comments.
This work has made use of archive data obtained with the Very Long Baseline Array. The National Radio Astronomy Observatory (NRAO VLBA) is a facility of the National Science Foundation operated under cooperative agreement by Associated Universities, Inc. 
This work made use of the Swinburne University of Technology software correlator \citep{2011PASP..123..275D}, developed as part of the Australian Major National Research Facilities Programme and operated under license.
This work has made use of observations obtained with the 14 m Mets\"ahovi Radio Observatory, a separate research institute of the Helsinki University of Technology. The Mets\"ahovi team acknowledges the support from the Academy of Finland to our observing projects (numbers 212656, 210338, 121148, and others). \\

\clearpage

\clearpage

\begin{deluxetable}{cccr}
\tablewidth{0pt}
\tablecaption{Observational Information.
\label{tab:obsinfo}}
\tablehead{\colhead{Array} & \colhead{Epoch} & \colhead{Band \tablenotemark{a}} & \colhead{Time On \tablenotemark{b}}\\
 & & & \colhead{Source (min)}}
\startdata
VLBA	&	2002 Jan 7	&	Q	&	32	\\
VLBA	&	2002 Apr 5	&	Q	&	704	\\
VLBA	&	2003 Jan 12	&	Q	&	25	\\
VLBA	&	2003 Nov 20	&	Q	&	6	\\
VLBA	&	2004 Oct 2	&	Q	&	6	\\
VLBA	&	2005 Jan 25	&	Q	&	7	\\
VLBA	&	2005 Oct 4	&	Q	&	6	\\
VLBA	&	2006 Jan 7	&	Q	&	9	\\
VLBA	&	2006 Mar 22	&	Q	&	9	\\
VLBA	&	2006 Apr 7	&	Q	&	6	\\
VLBA	&	2006 May 08	&	Q	&	6	\\
VLBA	&	2006 May 26	&	Q	&	6	\\
VLBA	&	2006 Jul 26	&	Q	&	9	\\
VLBA	&	2006 Sep 1	&	Q	&	6	\\
VLBA	&	2006 Sep 22	&	Q	&	6	\\
VLBA	&	2007 Jan 4	&	Q	&	9	\\
VLBA	&	2007 May 24	&	Q	&	9	\\
VLBA	&	2007 Jun 15	&	Q	&	6	\\
VLBA	&	2007 Aug 7	&	Q	&	5	\\
VLBA	&	2007 Nov 2	&	Q	&	6	\\
VLBA	&	2008 Apr 3	&	Q	&	6	\\
VLBA	&	2008 Apr 16	&	Q	&	15	\\
VLBA	&	2008 May 8	&	Q	&	5	\\
VLBA	&	2008 May 23	&	Q	&	6	\\
VLBA	&	2008 Jun 31	&	Q	&	6	\\
VLBA	&	2008 Aug 27	&	Q	&	6	\\
VLBA	&	2008 Nov 1	&	Q	&	6	\\
VLBA	&	2008 Nov 27	&	K,Q	&	6	\\
\enddata
\tablenotetext{a}{Band K = 22~GHz, Q = 43~GHz.}
\tablenotetext{b}{Total on source time in min.}
\end{deluxetable}

\clearpage

\begin{deluxetable}{cccc}
\tablewidth{0pt}
\tablecaption{Dispersions of C3 positions.
\label{tab:pairs}}
\tablehead{
\colhead{Pairs\tablenotemark{a}} & \colhead{$\theta _B^{mean}$~\tablenotemark{b}} & \colhead{$\Delta$ RA~\tablenotemark{c}} & \colhead{$\Delta$ Dec~\tablenotemark{c}}\\
\colhead{} & \colhead{(mas)} & \colhead{(beam)} & \colhead{(beam)}}
\startdata
2004 Oct  2 / 2004 Oct  3 &	0.31 	&	-0.01 	&	-0.01 	\\ 
2005 Oct  4 / 2005 Oct 12 &	0.33 	&	-0.20 	&	 0.09 	\\ 
2005 Oct 12 / 2005 Oct 18 &	0.33 	&	 0.16 	&	 0.06 	\\ 
2006 Jan  6 / 2006 Jan  7 &	0.27 	&	-0.04 	&	-0.03 	\\ 
2006 Mar 22 / 2006 Mar 23 &	0.27 	&	 0.07 	&	-0.11 	\\ 
2006 Apr  7 / 2006 Apr 15 &	0.32 	&	 0.26 	&	 0.07 	\\ 
2006 May  8 / 2006 May 12 &	0.34 	&	 0.06 	&	-0.19 	\\ 
2006 Sep 22 / 2006 Sep 28 &	0.28 	&	-0.01 	&	 0.05 	\\ 
2006 Dec 28 / 2007 Jan  4 &	0.30 	&	 0.09 	&	 0.01 	\\ 
2007 Aug  7 / 2007 Aug 17 &	0.33 	&	-0.22 	&	-0.14 	\\ 
2008 Nov  1 / 2008 Nov 11 &	0.36 	&	-0.17 	&	-0.14 	\\
\enddata
\tablenotetext{a}{Pairs of near epochs with their separation $\leq $ 10 days (for which the motions of components are negligible)}.
\tablenotetext{b}{Beam sizes averaged for the major - minor axis and two epochs.}
\tablenotetext{c}{Differences of relative right ascension and declination position of C3 between two epochs.}
\end{deluxetable}

\clearpage

\begin{deluxetable}{ccccc}

%\tablewidth{0pt}
\tablecaption{Epoch, rms, synthesized beam size, peak brightness, total CLEANed flux for all image.
\label{tab:VLBA_Q_sum}}
\tablehead{
\colhead{Epoch}	& \colhead{rms}	& \colhead{Beam \tablenotemark{a}}	&	\colhead{$I_{\rm peak}$ \tablenotemark{b}} & \colhead{$S$ \tablenotemark{c}} \\
\colhead{}	&	\colhead{(mJy beam$^{-1}$)}	& \colhead{(mas $\times$ mas, degree)} & \colhead{(Jy beam$^{-1}$)} & \colhead{(Jy)} 
}
\startdata
2002 Jan 7	&	8.5 	&	0.31 	$\times$	0.19 	,	-9.0 	&	1.07 	&	3.48 	$\pm$	0.35 	\\
2002 Apr 5	&	6.6 	&	0.25 	$\times$	0.14 	,	-33.7 	&	0.96 	&	3.27 	$\pm$	0.33 	\\
2003 Jan 12	&	10.6 	&	0.39 	$\times$	0.16 	,	3.1 	&	1.00 	&	4.07 	$\pm$	0.41 	\\
2003 Nov 20	&	23.7 	&	0.53 	$\times$	0.18 	,	33.6 	&	1.17 	&	4.29 	$\pm$	0.43 	\\
2004 Oct 2	&	13.0 	&	0.49 	$\times$	0.14 	,	-23.1 	&	1.59 	&	4.42 	$\pm$	0.44 	\\
2005 Jan 25	&	12.8 	&	0.29 	$\times$	0.17 	,	-22.2 	&	1.04 	&	4.10 	$\pm$	0.41 	\\
2005 Oct 4	&	18.8 	&	0.17 	$\times$	0.46 	,	-34.4 	&	1.49 	&	4.60 	$\pm$	0.46 	\\
2006 Jan 7	&	39.8 	&	0.27 	$\times$	0.19 	,	-5.7 	&	1.44 	&	4.40 	$\pm$	0.44 	\\
2006 Mar 22	&	17.3 	&	0.20 	$\times$	0.33 	,	-6.8 	&	1.18 	&	3.72 	$\pm$	0.37 	\\
2006 Apr 7	&	68.4 	&	0.52 	$\times$	0.18 	,	14.4 	&	1.60 	&	4.27 	$\pm$	0.43 	\\
2006 May 8	&	45.3 	&	0.26 	$\times$	0.43 	,	-23.0 	&	1.75 	&	5.07 	$\pm$	0.51 	\\
2006 May 26	&	13.4 	&	0.48 	$\times$	0.17 	,	-33.0 	&	1.30 	&	4.34 	$\pm$	0.43 	\\
2006 Jul 26	&	36.3 	&	0.44 	$\times$	0.23 	,	-4.2 	&	1.58 	&	4.20 	$\pm$	0.42 	\\
2006 Sep 1	&	15.5 	&	0.16 	$\times$	0.48 	,	-34.2 	&	1.32 	&	3.77 	$\pm$	0.38 	\\
2006 Sep 22	&	34.2 	&	0.25 	$\times$	0.19 	,	-13.0 	&	1.45 	&	4.94 	$\pm$	0.49 	\\
2007 Jan 4	&	38.7 	&	0.28 	$\times$	0.19 	,	-15.8 	&	1.34 	&	4.48 	$\pm$	0.45 	\\
2007 May 24	&	20.6 	&	0.22 	$\times$	0.19 	,	-11.7 	&	0.99 	&	3.93 	$\pm$	0.39 	\\
2007 Jun 15	&	39.4 	&	0.26 	$\times$	0.55 	,	14.5 	&	2.04 	&	4.37 	$\pm$	0.44 	\\
2007 Aug 7	&	45.0 	&	0.51 	$\times$	0.23 	,	16.9 	&	1.73 	&	5.43 	$\pm$	0.54 	\\
2007 Nov 2	&	39.2 	&	0.18 	$\times$	0.53 	,	-1.9 	&	2.16 	&	4.62 	$\pm$	0.46 	\\
2008 Apr 3	&	33.7 	&	0.14 	$\times$	0.47 	,	-22.5 	&	1.85 	&	5.82 	$\pm$	0.58 	\\
2008 Apr 16	&	14.4 	&	0.45 	$\times$	0.19 	,	-15.3 	&	1.93 	&	4.93 	$\pm$	0.49 	\\
2008 May 8	&	41.0 	&	0.49 	$\times$	0.14 	,	-11.1 	&	2.00 	&	5.97 	$\pm$	0.60 	\\
2008 May 23	&	37.8 	&	0.46 	$\times$	0.15 	,	20.9 	&	1.43 	&	4.36 	$\pm$	0.44 	\\
2008 Jul 31	&	46.5 	&	0.51 	$\times$	0.15 	,	-12.2 	&	2.02 	&	6.33 	$\pm$	0.63 	\\
2008 Aug 27	&	34.3 	&	0.54 	$\times$	0.20 	,	27.3 	&	2.46 	&	6.89 	$\pm$	0.69 	\\
2008 Nov 1	&	57.1 	&	0.53 	$\times$	0.18 	,	-1.0 	&	2.85 	&	9.19 	$\pm$	0.92 	\\
2008 Nov 27	&	61.3 	&	0.45 	$\times$	0.14 	,	7.6 	&	2.22 	&	8.90 	$\pm$	0.89 	\\
\enddata
\tablenotetext{a}{Major axis, minor axis, and position angle of synthesized beam.}
\tablenotetext{b}{Peak brightness for each image.}
\tablenotetext{c}{Total CLEANed Flux and its error for each image. We assumed that the flux calibration error is 10\% of flux density.}
\label{tab:VLBA_Q}
\end{deluxetable}

\clearpage

\begin{deluxetable}{ccc}																	
\tablewidth{0pt}
\tablecaption{model type and flux of fitted C1 component.\label{tab:model_C1}}
\tablehead{
\colhead{Epoch}	& \colhead{Model\tablenotemark{a}}	& \colhead{$S_{\rm mod}$ \tablenotemark{b}}	\\
\colhead{}		&									& \colhead{(Jy)} 
}																			
\startdata
2002 Jan 7	2002	&	e	&	2.13	$\pm$	0.22	\\
2002 Apr 5	2002	&	e	&	1.16	$\pm$	0.12	\\
2003 Jan 12	2003	&	e	&	2.04	$\pm$	0.21	\\
2003 Nov 20	2003	&	e	&	2.68	$\pm$	0.29	\\
2004 Oct 2	2004	&	e	&	2.15	$\pm$	0.22	\\
2005 Jan 25	2005	&	e	&	2.27	$\pm$	0.24	\\
2005 Oct 4	2005	&	c	&	1.93	$\pm$	0.22	\\
2006 Jan 7	2006	&	c	&	2.13	$\pm$	0.22	\\
2006 Mar 22	2006	&	e	&	1.59	$\pm$	0.17	\\
2006 Apr 7	2006	&	e	&	2.61	$\pm$	0.30	\\
2006 May 8	2006	&	c	&	2.09	$\pm$	0.22	\\
2006 May 26	2006	&	c	&	1.34	$\pm$	0.13	\\
2006 Jul 26	2006	&	c	&	1.64	$\pm$	0.17	\\
2006 Sep 1	2006	&	e	&	1.74	$\pm$	0.18	\\
2006 Sep 22	2006	&	c	&	1.77	$\pm$	0.18	\\
2007 Jan 4	2007	&	c	&	2.01	$\pm$	0.22	\\
2007 May 24	2007	&	e	&	1.69	$\pm$	0.18	\\
2007 Jun 15	2007	&	c	&	1.97	$\pm$	0.20	\\
2007 Aug 7	2007	&	c	&	1.72	$\pm$	0.18	\\
2007 Nov 2	2007	&	c	&	2.62	$\pm$	0.28	\\
2008 Apr 3	2008	&	c	&	2.22	$\pm$	0.23	\\
2008 Apr 16	2008	&	c	&	1.89	$\pm$	0.19	\\
2008 May 8	2008	&	c	&	2.46	$\pm$	0.25	\\
2008 May 23	2008	&	e	&	2.15	$\pm$	0.23	\\
2008 Jan 31	2008	&	c	&	2.67	$\pm$	0.28	\\
2008 Aug 27	2008	&	e	&	3.37	$\pm$	0.35	\\
2008 Nov 1	2008	&	e	&	4.36	$\pm$	0.47	\\
2008 Nov 27	2008	&	c	&	3.49	$\pm$	0.37	\\
\enddata
\tablenotetext{a}{Shape of component models. 'e' is elliptical Gaussian, 'c' is circle Gaussian function.}
\tablenotetext{b}{Flux and its error for models. Error is estimated as the addition of calibration error (typically 10\% of component) and image rms of each epoch.}
\end{deluxetable}

\clearpage

\begin{deluxetable}{ccccc}																				
\tablewidth{0pt}																			
\tablecaption{Relative position from C1, model type, and flux of fitted C2 component.\label{tab:model_C2}}
\tablehead{
\colhead{Epoch}	& \colhead{Relative R.A.\tablenotemark{a}} & \colhead{Relative Dec.\tablenotemark{a}} & \colhead{Model\tablenotemark{b}}	&	
\colhead{$S_{\rm mod}$ \tablenotemark{c}}	\\
\colhead{}	&	\colhead{(mas)}	& \colhead{(mas)} &	& \colhead{(Jy)} 
}			
\startdata																		
2002 Jan 7	2002	&	-0.55	&	-0.78	&	e	&	1.19	$\pm$	0.14	\\
2002 Apr 5	2002	&	-0.66	&	-0.88	&	e	&	0.54	$\pm$	0.07	\\
2003 Jan 12	2003	&	-0.65	&	-0.77	&	e	&	0.88	$\pm$	0.10	\\
2003 Nov 20	2003	&	-0.36	&	-0.76	&	e	&	0.59	$\pm$	0.09	\\
2004 Oct 2	2004	&	-0.70	&	-0.93	&	e	&	0.96	$\pm$	0.13	\\
2005 Jan 25	2005	&	-0.81	&	-0.93	&	c	&	0.42	$\pm$	0.06	\\
2005 Oct 4	2005	&	-0.53	&	-0.95	&	e	&	1.32	$\pm$	0.15	\\
2006 Jan 7	2006	&	-0.62	&	-1.12	&	e	&	0.84	$\pm$	0.08	\\
2006 Mar 22	2006	&	-0.64	&	-1.13	&	e	&	0.81	$\pm$	0.10	\\
2006 Apr 7	2006	&	-0.82	&	-1.25	&	e	&	0.70	$\pm$	0.11	\\
2006 May 8	2006	&	-0.57	&	-1.16	&	e	&	1.24	$\pm$	0.16	\\
2006 May 26	2006	&	-0.53	&	-1.11	&	e	&	1.44	$\pm$	0.18	\\
2006 Jul 26	2006	&	-0.73	&	-1.07	&	e	&	0.94	$\pm$	0.14	\\
2006 Sep 1	2006	&	-0.55	&	-1.15	&	e	&	1.19	$\pm$	0.16	\\
2006 Sep 22	2006	&	-0.56	&	-1.13	&	e	&	1.35	$\pm$	0.20	\\
2007 Jan 4	2007	&	-0.65	&	-1.19	&	e	&	0.66	$\pm$	0.11	\\
2007 May 24	2007	&	-0.57	&	-1.40	&	e	&	0.43	$\pm$	0.06	\\
2007 Jun 15	2007	&	-0.74	&	-1.30	&	e	&	0.61	$\pm$	0.08	\\
2007 Aug 7	2007	&	-0.96	&	-1.27	&	e	&	0.63	$\pm$	0.06	\\
2007 Nov 2	2007	&	-0.69	&	-1.34	&	c	&	0.61	$\pm$	0.10	\\
2008 Apr 3	2008	&	-0.63	&	-1.24	&	e	&	0.82	$\pm$	0.20	\\
2008 Apr 16	2008	&	-0.57	&	-1.22	&	c	&	0.82	$\pm$	0.13	\\
2008 May 8	2008	&	-0.74	&	-1.11	&	c	&	0.53	$\pm$	0.16	\\
2008 May 23	2008	&	-0.60	&	-1.16	&	c	&	0.57	$\pm$	0.13	\\
2008 Jan 31	2008	&	-0.83	&	-1.07	&	c	&	0.35	$\pm$	0.04	\\
2008 Aug 27	2008	&	-0.35	&	-1.24	&	d	&	0.51	$\pm$	0.05	\\
2008 Nov 1	2008	&	-0.74	&	-1.49	&	d	&	0.37	$\pm$	0.04	\\
2008 Nov 27	2008	&	-0.71	&	-1.47	&	d	&	0.36	$\pm$	0.04	\\
\enddata																								
\tablenotetext{a}{Relative R.A. and Dec. angular distances between C1 and C2.}
\tablenotetext{b}{Shape of component models. 'e' is elliptical Gaussian, 'c' is circle Gaussian, and 'd' is delta function.}
\tablenotetext{c}{Flux and its error for models. Error is estimated as the addition of calibration error (typically 10\% of component) and image rms of each epoch.}
\end{deluxetable}

\clearpage

\begin{deluxetable}{cccccc}																				
\tablewidth{0pt}																			
\tablecaption{Relative position from C1, model type, size, and flux of fitted C3 component.\label{tab:model_C3}}																			
\tablehead{
\colhead{Epoch}	& \colhead{Relative R.A.\tablenotemark{a}} & \colhead{Relative Dec.\tablenotemark{a}} & \colhead{Model\tablenotemark{b}}	&	
\colhead{$\theta _{\rm maj} \times \theta _{\rm min}$, P.A. \tablenotemark{c}}	&	\colhead{$S_{\rm mod}$ \tablenotemark{d}}	\\
\colhead{}	&	\colhead{(mas)}	& \colhead{(mas)} &	& \colhead{(mas $\times$ mas, deg)} & \colhead{(Jy)} 
}																	
\startdata																			
2002 Jan 7	2002	&	\nodata	&	\nodata	&	\nodata	&	\nodata								&		\nodata		\\
2002 Apr 5	2002	&	\nodata	&	\nodata	&	\nodata	&	\nodata								&		\nodata		\\
2003 Jan 12	2003	&	\nodata	&	\nodata	&	\nodata	&	\nodata								&		\nodata		\\
2003 Nov 20	2003	&	0.04	&	-0.12	&	c	&	0.13	$\times$	0.13	,	0.0		&	0.63	$\pm$	0.07	\\
2004 Oct 2	2004	&	0.09	&	-0.25	&	e	&	0.27	$\times$	0.00	,	66.3	&	1.14	$\pm$	0.11	\\
2005 Jan 25	2005	&	0.12	&	-0.24	&	c	&	0.05	$\times$	0.05	,	0.0		&	0.32	$\pm$	0.03	\\
2005 Oct 4	2005	&	0.13	&	-0.46	&	e	&	0.23	$\times$	0.09	,	27.2	&	1.34	$\pm$	0.14	\\
2006 Jan 7	2006	&	0.14	&	-0.47	&	c	&	0.13	$\times$	0.13	,	0.0		&	0.67	$\pm$	0.07	\\
2006 Mar 22	2006	&	0.13	&	-0.50	&	c	&	0.23	$\times$	0.23	,	0.0		&	1.01	$\pm$	0.11	\\
2006 Apr 7	2006	&	0.14	&	-0.48	&	c	&	0.06	$\times$	0.06	,	0.0		&	0.81	$\pm$	0.08	\\
2006 May 8	2006	&	0.13	&	-0.50	&	c	&	0.23	$\times$	0.23	,	0.0		&	1.72	$\pm$	0.19	\\
2006 May 26	2006	&	0.12	&	-0.52	&	e	&	0.43	$\times$	0.23	,	1.9		&	1.71	$\pm$	0.19	\\
2006 Jul 26	2006	&	0.03	&	-0.45	&	c	&	0.20	$\times$	0.20	,	0.0		&	1.01	$\pm$	0.12	\\
2006 Sep 1	2006	&	0.12	&	-0.54	&	c	&	0.32	$\times$	0.32	,	0.0		&	1.18	$\pm$	0.14	\\
2006 Sep 22	2006	&	0.11	&	-0.55	&	c	&	0.29	$\times$	0.29	,	0.0		&	1.58	$\pm$	0.19	\\
2007 Jan 4	2007	&	0.04	&	-0.59	&	e	&	0.45	$\times$	0.21	,	-61.5	&	1.37	$\pm$	0.21	\\
2007 May 24	2007	&	0.15	&	-0.84	&	c	&	0.22	$\times$	0.22	,	0.0		&	0.88	$\pm$	0.11	\\
2007 Jun 15	2007	&	0.17	&	-0.80	&	e	&	0.00	$\times$	0.00	,	0.0		&	0.91	$\pm$	0.09	\\
2007 Aug 7	2007	&	0.13	&	-0.95	&	e	&	0.55	$\times$	0.29	,	17.7	&	1.89	$\pm$	0.25	\\
2007 Nov 2	2007	&	0.06	&	-0.98	&	c	&	0.43	$\times$	0.28	,	45.0	&	1.74	$\pm$	0.27	\\
2008 Apr 3	2008	&	0.05	&	-1.06	&	e	&	0.34	$\times$	0.27	,	-43.1	&	1.55	$\pm$	0.20	\\
2008 Apr 16	2008	&	0.10	&	-1.09	&	e	&	0.44	$\times$	0.26	,	25.7	&	1.62	$\pm$	0.18	\\
2008 May 8	2008	&	0.09	&	-1.08	&	e	&	0.34	$\times$	0.15	,	68.1	&	1.61	$\pm$	0.19	\\
2008 May 23	2008	&	0.07	&	-1.07	&	e	&	0.58	$\times$	0.33	,	16.2	&	2.04	$\pm$	0.31	\\
2008 Jan 31	2008	&	0.06	&	-1.11	&	e	&	0.72	$\times$	0.34	,	47.0	&	2.91	$\pm$	0.44	\\
2008 Aug 27	2008	&	0.07	&	-1.11	&	e	&	0.65	$\times$	0.27	,	8.5		&	2.67	$\pm$	0.32	\\
2008 Nov 1	2008	&	0.05	&	-1.23	&	e	&	0.58	$\times$	0.44	,	45.0	&	3.76	$\pm$	0.53	\\
2008 Nov 27	2008	&	0.08	&	-1.27	&	e	&	0.67	$\times$	0.34	,	43.1	&	3.57	$\pm$	0.58	\\
\enddata																	
\tablenotetext{a}{Relative R.A. and Dec. angular distances between C1 and C3.}
\tablenotetext{b}{Shape of component models. 'e' is elliptical Gaussian, 'c' is circle Gaussian function.}
\tablenotetext{c}{Major axis, minor axis, and position angle of models. $\theta _{\rm maj} $ and $\theta _{\rm min}$ are not defined for delta models. P.A. is not defined for circle and delta models.}
\tablenotetext{d}{Flux and its error for models. Error is estimated as the addition of calibration error (typically 10\% of component) and image rms of each epoch.}
\end{deluxetable}
\clearpage

%% Tables may also be prepared as separate files. See the accompanying
%% sample file table.tex for an example of an external table file.
%% To include an external file in your main document, use the \input
%% command. Uncomment the line below to include table.tex in this
%% sample file. (Note that you will need to comment out the \documentclass,
%% \begin{document}, and \end{document} commands from table.tex if you want
%% to include it in this document.)


%% The following command ends your manuscript. LaTeX will ignore any text


\begin{thebibliography}{}

\bibitem[Abdo et al.(2009)]{2009ApJ...699...31A} Abdo, A.~A., et al.\ 2009, \apj, 699, 31 

\bibitem[Abdo et al.(2010)]{2010ApJ...720..912A} Abdo, A.~A., et al.\ 2010, \apj, 720, 912 

\bibitem[Asada et al.(2006)]{2006PASJ...58..261A} Asada, K., Kameno, S., Shen, Z.-Q., Horiuchi, S., Gabuzda, D.~C., \& Inoue, M.\ 2006, \pasj, 58, 261 

\bibitem[Barnes(1994)]{1994saes.book.....B} Barnes, J.~W.,\ 1994, ``Statistical Analysis for Engineers and Scientists'' New York, McGraw-Hill, Inc., 

\bibitem[Deller et al.(2011)]{2011PASP..123..275D} Deller, A.~T., et al.\ 2011, \pasp, 123, 275 

\bibitem[Georganopoulos \& Kazanas(2003)]{2003ApJ...594L..27G} Georganopoulos, M., \& Kazanas, D.\ 2003, \apjl, 594, L27 

\bibitem[Ghisellini et al.(2005)]{2005A&A...432..401G} Ghisellini, G., Tavecchio, F., \& Chiaberge, M.\ 2005, \aap, 432, 401 

\bibitem[Hartman et al.(2008)]{2008ApJ...688..852H} Hartman, R.~C., Kadler, M., \& Tueller, J.\ 2008, \apj, 688, 852 

\bibitem[Kataoka et al.(2000)]{2000ApJ...528..243K} Kataoka, J., Takahashi, T., Makino, F., et al.\ 2000, \apj, 528, 243 

\bibitem[Kino et al.(2002)]{2002ApJ...564...97K} Kino, M., Takahara, F., \& Kusunose, M.\ 2002, \apj, 564, 97 

\bibitem[Le{\'o}n-Tavares et al.(2011)]{2011A&A...532A.146L} Le{\'o}n-Tavares, J., Valtaoja, E., Tornikoski, M., L{\"a}hteenm{\"a}ki, A., \& Nieppola, E.\ 2011, \aap, 532, A146 

\bibitem[Lister et al.(2009)]{2009AJ....138.1874L} Lister, M.~L., et al.\ 2009, \aj, 138, 1874  

\bibitem[Liuzzo et al.(2010)]{2010A&A...516A...1L} Liuzzo, E., Giovannini, G., Giroletti, M., \& Taylor, G.~B.\ 2010, \aap, 516, A1 

\bibitem[Mukherjee et al.(2002)]{2002ApJ...574..693M} Mukherjee, R., Halpern, J., Mirabal, N., \& Gotthelf, E.~V.\ 2002, \apj, 574, 693 

\bibitem[Nagai et al.(2010)]{2010PASJ...62L..11N} Nagai, H., et al.\ 2010, \pasj, 62, L11 ({\bf Paper I})

\bibitem[Reimer et al.(2003)]{2003ApJ...588..155R} Reimer, O., Pohl, M., Sreekumar, P., \& Mattox, J.~R.\ 2003, \apj, 588, 155 

\bibitem[Shepherd et al.(1994)]{1994BAAS...26..987S} Shepherd, M.~C., Pearson, T.~J., \& Taylor, G.~B.\ 1994, \baas, 26, 987 

\bibitem[Sreekumar et al.(1999)]{1999APh....11..221S} Sreekumar, P., Bertsch, D.~L., Hartman, R.~C., Nolan, P.~L., \& Thompson, D.~J.\ 1999, Astroparticle Physics, 11, 221 

\bibitem[Ter\"{a}sranta et al.(1998)]{1998A&AS..132..305T} Ter\"{a}sranta, H., et al.\ 1998, \aaps, 132, 305 

\bibitem[Vermeulen et al.(1994)]{1994ApJ...430L..41V} Vermeulen, R.~C., Readhead, A.~C.~S., \& Backer, D.~C.\ 1994, \apjl, 430, L41 

\bibitem[Walker et al.(1994)]{1994ApJ...430L..45W} Walker, R.~C., Romney, J.~D., \& Benson, J.~M.\ 1994, \apjl, 430, L45 

\bibitem[Walker et al.(2000)]{2000ApJ...530..233W} Walker, R.~C., Dhawan, V., Romney, J.~D., Kellermann, K.~I., \& Vermeulen, R.~C.\ 2000, \apj, 530, 233 

\end{thebibliography}
\end{document}